\documentclass[conference,das]{IEEEtran}
\IEEEoverridecommandlockouts

\usepackage{cite}
\usepackage{amsmath,amssymb,amsfonts}
\usepackage{algorithmic}
\usepackage{graphicx}
\usepackage{textcomp}
\usepackage{fancyhdr}
\usepackage{xcolor}
\usepackage{url}
\def\BibTeX{{\rm B\kern-.05em{\sc i\kern-.025em b}\kern-.08em
    T\kern-.1667em\lower.7ex\hbox{E}\kern-.125emX}}

\fancypagestyle{firstpage}
{
   \fancyhf{}
   \fancyhead[R]{\small LLNL-CONF-2004165}
   \fancyfoot[C]{\thepage}
}
\fancypagestyle{otherpage}
{
   \fancyhf{}
   \fancyfoot[C]{\thepage}
}
\begin{document}

\newcommand\andre[1]{{\color{blue}[ANDR\'{E}: #1]}}
\newcommand\colin[1]{{\color{red}[COLIN: #1]}}
\newcommand\sungwoo[1]{{\color{magenta}[SUNGWOO: #1]}}
\newcommand\asmcomment[1]{{\color{orange}[AARON: #1]}}
\newcommand\asm[1]{{\color{orange}{#1}}}
\newcommand\kate[1]{{\color{cyan}[KATE: #1]}}
\newcommand\pavlos[1]{{\color{green}[PAVLOS: #1]}}
\newcommand\frl[1]{{\color{brown}[FERNANDO: #1]}}
\newcommand\sfk[1]{{\color{olive}[SK: #1]}}
\newcommand\all[1]{{\color{green}[ALL: #1]}}
\newcommand\fade[1]{{\color{lightgray}{#1}}}
\newcommand\del[1]{}

\newcommand\Fig[1]{Fig.~\ref{fig:#1}}
\newcommand\eq[1]{Eq.~\ref{eq:#1}}
\newcommand\sect[1]{Sect.~\ref{sec:#1}}
\newcommand\refcite[1]{Ref.~\cite{#1}}
\newcommand\refscite[1]{Refs.~\cite{#1}}
\newcommand{\qudalaph}{quda\_laph}
\newcommand{\Qudalaph}{Quda\_laph}
\newcommand{\quda}{QUDA}
\newcommand\authorhide[1]{\author{#1}}
\newcommand\hide[1]{#1}

\newcommand{\su}{$\sim$240}
%

\title{{\bf \large 
Using Exascale Computing to Explain the Delicate Balance
of Nuclear Forces in the Universe}\\[-.25em]
{\large \emph{(The unseen harmony is better than the visible --
Heraclitus
500BC)}} 
\\[-.25em]
{\large Entry to the 2025 Gordon Bell Prize competition; April 14, 2025}
\vspace{-0.5em}
}

\authorhide{
\IEEEauthorblockN{
M.A. Clark\IEEEauthorrefmark{1},
A. Hanlon\IEEEauthorrefmark{2},
D. Howarth \IEEEauthorrefmark{3},
B. Joo \IEEEauthorrefmark{1},
S. Krieg \IEEEauthorrefmark{4}\IEEEauthorrefmark{11},
D. McDougall \IEEEauthorrefmark{5},
A. Meyer \IEEEauthorrefmark{6}\IEEEauthorrefmark{7},
\\
H. Monge-Camacho \IEEEauthorrefmark{8},
C. Morningstar \IEEEauthorrefmark{9},
S. Park \IEEEauthorrefmark{6}\IEEEauthorrefmark{7},
F. Romero-L\'opez \IEEEauthorrefmark{10},
P. M. Vranas \IEEEauthorrefmark{6}\IEEEauthorrefmark{7},
A. Walker-Loud \IEEEauthorrefmark{7}\IEEEauthorrefmark{6}
}
\smallskip
\IEEEauthorblockA{\IEEEauthorrefmark{1}
	NVIDIA Corporation, 
	Santa Clara, California, 95051, USA}
\IEEEauthorblockA{\IEEEauthorrefmark{2}
	Department of Physics, Kent State University, 
    Kent, OH 44242, USA}
\IEEEauthorblockA{\IEEEauthorrefmark{3}
    California Institute of Technology, Pasadena, California, 91125, USA}
\IEEEauthorblockA{\IEEEauthorrefmark{4}
    J\"ulich Supercomputing Centre and CASA, 
    Forschungszentrum Jülich, 
    52425 Jülich, Germany}      
\IEEEauthorblockA{\IEEEauthorrefmark{5}
    Advanced Micro Devices, Inc.,
    Santa Clara, California, 95054, USA}      
\IEEEauthorblockA{\IEEEauthorrefmark{6}
	Physical Sciences Directorate,
	Lawrence Livermore National Laboratory,
	Livermore, California 94550, USA}
\IEEEauthorblockA{\IEEEauthorrefmark{7}
	Nuclear Science Division,
	Lawrence Berkeley National Laboratory,
	Berkeley, CA 94720, USA}
\IEEEauthorblockA{\IEEEauthorrefmark{8}
    National Center For Computational Sciences,
    Oak Ridge National Laboratory, 
    Oak Ridge, Tennessee, 37831, USA
}
\IEEEauthorblockA{\IEEEauthorrefmark{9}
	Department of Physics, 
	Carnegie Mellon University, 
	Pittsburgh, PA 15213, USA}
\IEEEauthorblockA{\IEEEauthorrefmark{10}
	Albert Einstein Center, Institute for Theoretical Physics, University of Bern, 3012 Bern, Switzerland
}
\IEEEauthorblockA{\IEEEauthorrefmark{11}
    HISKP,
    University of Bonn,
    Regina-Pacis-Weg 3, 
    53113 Bonn, Germany}   
}

\maketitle
\thispagestyle{firstpage}

\begin{abstract}
The vast majority of visible matter in our universe comes from protons and neutrons (the nucleons). Nucleon interactions are fundamental to how the universe developed after the Big Bang and govern all nuclear phenomena. 
The subtle balance in how two nucleons interact shapes the universe’s hydrogen content that is central to our existence.
Our objective is to compute the interaction strength while varying the parameters of nature to understand how delicate this balance is.
We developed a new code using sophisticated physics algorithms and a highly optimized library for simulations on CPU-GPU parallel architectures. It has excellent weak scaling and impressive linear scaling for a fixed problem size with increasing number of nodes up to El Capitan's full $\sim$11,000 nodes. On Alps, El Capitan, Frontier, Jupiter, and Perlmutter supercomputers we achieve a maximum disruptive speed-up of \su{} times the previous state-of-the-art, signaling a new era of supercomputing.
\looseness-1
\end{abstract}

\section{\label{sec:just} Justification for ACM Gordon Bell Prize}
We address a fundamental mystery of nature. Our algorithms and code exhibit excellent weak scaling and impressive linear scaling with increasing number of nodes up to El Capitan's $\sim 11,000$ nodes. It performs excellently on Alps, El Capitan, Frontier, Jupiter, and Perlmutter supercomputers with a maximum disruptive speed-up of \su.

\section{\label{sec:perf_attributes} Performance Attributes}

\begin{table}[h]
\centering
\begin{tabular}{|c|c|}
\hline
Attribute                    & Value \\
\hline
Category of achievement & scalability, time to solution \\
\hline
method                    & explicit \\
\hline
reporting                   & whole application including I/O \\
\hline
precision                   & mixed-precision \\
\hline
system scale              & full-scale system \\
\hline
measurement method & timers \\
\hline
\end{tabular}
\vspace{.1in}
\caption{Performance Attributes}
\label{tab:attributes}
\end{table}

\section{\label{sec:overview} Overview }
Our existence depends upon the composition of visible matter in the Universe which is comprised of roughly 74\% hydrogen (H) and 24-25\% helium-4 ($^4$He) while the rest of the nuclear elements make up a mere 1-2\% of this visible mass.
This large abundance of H is important because it allows the universe to have many H-burning stars, like our Sun, which are very stable over cosmic time scales.
H-burning stars support themselves from gravitational collapse by fusing two protons into a deuteron (D), the lightest nucleus formed from one proton and one neutron, through the conversion of a proton into a neutron via the {\it weak nuclear force}.
In the core of the Sun, the average lifetime of a proton before fusing into a D, is roughly 9 billion years, providing a stable solar environment to support the development of intelligent life on Earth.

In contrast, the helium burning stage of stars is a mere 100 million years. Stages of burning that synthesize successively heavier elements, such as carbon, nitrogen and oxygen, continue to rapidly decrease in duration until the silicon burning stage that occurs in heavier stars, which lasts only about a day before the star collapses and explodes within minutes.  Without an abundance of H, the Universe would be depleted of stars like our Sun, and thus we would not be here to contemplate our existence.

At the same time, the abundance of H versus $^4$He and other elements is delicately balanced, as we will now discuss.  To understand these abundances, we turn to the primordial universe in
an epoch that occurred within the first few minutes after the Big Bang during which $^4$He, beryllium, lithium and a few other light elements were synthesized from the initial protons and neutrons.
During this time, the light nuclear elements formed through a process known as Big Bang Nucleosynthesis (BBN)~\cite{Fields:2019pfx}.

The neutron, being slightly heavier than a proton, decays to a proton, electron and electron-anti-neutrino through the same {\it weak force} mentioned above, with a lifetime of about 15 minutes.  At the beginning of BBN, the universe contained roughly equal numbers of neutrons and protons (the nucleons).  This is because the {\it weak} reactions, which convert neutrons to protons, and vice versa, were happening sufficiently fast compared to the expansion of the universe, that the states were held essentially in thermodynamic equilibrium such that the ratio of neutrons ($n$) to protons ($p$) was simply set by the mass splitting divided by the temperature of the universe ($T$)
\begin{equation*}
\frac{X_n}{X_p} \simeq e^{-\frac{M_n - M_p}{T}}\, .
\end{equation*}

As the universe cooled off while expanding, such that $T\ll M_n-M_p$, this relation suggests there would be an exponentially small number of neutrons compared to protons.
Yet, from the post BBN abundances inferred from observations of the cosmos, the 74\% H ($p$) and 24-25\% $^4$He ($ppnn$), we know that there was roughly one neutron for every 7 protons.  A natural question to ask is,
{\it Why does the universe contain any neutrons at all?}

The answer to this question is the existence of the deuteron.
Once a neutron becomes bound in a deuterium nucleus, it is no longer energetically favorable for it to decay to a proton, and thus the mass fraction of neutrons to protons in the universe today is set by the existence of the deuteron and its binding energy, $B_D$.
Early deuterons did not persist for long because the primordial universe was mostly filled with radiation, such that there were roughly $10^9$ photons for every proton or neutron.
Thus, early in BBN, as soon as a deuteron was formed, it was immediately dissociated by the hot photon gas until the universe cooled enough that the probability of a deuteron colliding with an energetic photon reduced below the deuteron formation rate.  Nuclei heavier than D, such as $^4$He, could not appreciably synthesize during this time, which is known as the {\it deuterium bottleneck}.

After the universe expanded and cooled to a temperature of $T\approx0.1$~MeV, deuterium began to survive and then the universe rapidly synthesized tritium ($pnn$), $^3$He ($ppn$), $^4$He and a few other light elements: heavier elements such as carbon and oxygen were synthesized much later inside stars where the density and temperatures are higher. 

The deuteron is special in its own right because of its small $B_D\approx2$ million electron volts (MeV), or 1 MeV per nucleon which represents 0.1\% of its total mass.
The binding energy per nucleon of a typical nucleus is $\approx8$~MeV, or an order of magnitude larger.
For the deuteron, this represents what is referred to as a finely tuned system: if the strength of the nuclear force were to change by a small amount, we'd expect the deuteron to ``relax'' to a more ``natural'' state with a binding energy more typical of the other nuclei.
However, if the deuteron had a more natural binding energy, the universe would contain far less H and much more $^4$He:
if $B_D$ were larger (more natural), the temperature at which deuterons would form would be much higher, at an earlier time when there were many more neutrons, and thus those neutrons would be captured into stable $^4$He leaving a primordial Universe that was deplete of H, or at least with much less H-abundance.
There would therefore be many fewer stable H-burning stars like our Sun and thus we would not be here to consider questions of our origin.

This raises the question, can we understand how $B_D$ arises from the fundamental theory of particle physics known as the Standard Model?  For example, if some of the constants of the Standard Model were slightly different, would the deuteron still bind?  Would it have a more natural binding energy?  How different would the composition of the universe be?
Answering such questions can place stringent constraints on the possible variation of these constants in the early universe.

The major challenge to answer such questions is that the {\it strong nuclear force}, which binds protons and neutrons into nuclei, is notoriously difficult to understand from first principles.  While nucleons are the degrees of freedom of nuclear physics, they are not fundamental.  Rather, nucleons are composite states of quarks and gluons, which are fundamental degrees of freedom of Quantum Chromodynamics (QCD).  The massless gluons and nearly massless quarks are confined into the nucleons we observe in nature, with a mass of 1 billion electron-volts (GeV).  Around 95\% of the mass of the nucleons is the result of these QCD interactions, a remarkable feature of QCD, with only 5\% arising from the mass of the quarks~\cite{Durr:2015dna}.
QCD is one of the three fundamental theories that make up the Standard Model, which is our most precisely verified theory of nature.

After confinement, there is a relatively small residual interaction that binds the nucleons into atomic nuclei through two (NN), three (NNN) and higher order nucleon forces. 
While the $\approx8$~MeV binding energy per nucleon of atomic nuclei is two orders of magnitude smaller than the
nucleon mass, it is still very strong compared to the other known forces.
In conjunction with the other two forces of the Standard Model,
electromagnetism and the weak nuclear force,
the entirety of the rich field of nuclear physics all emerges from QCD:
from the forces binding protons and neutrons into the nuclear landscape,
to the fusion and fission reactions between nuclei,
and to the unknown state of matter at the cores of neutron stars.

In order to quantitatively understand $B_D$ from QCD, as well as the properties and reactions of nuclei, we must be able to predict the NN and NNN interactions from the underlying quark and gluon degrees of freedom.  The only way to obtain these robust predictions is by using numerical simulations on a discrete mesh, called the lattice, with a technique known as lattice QCD (LQCD).
LQCD calculations require state-of-the-art high-performance computing (HPC) to carry out large scale integrals using Monte Carlo methods.
The advent of GPUs more than a decade ago, along with algorithmic developments, disruptively improved the application of LQCD to many simpler quantities such as the neutron-proton mass splitting and a quantity related to the neutron lifetime, that has been computed with sub-percent precision~\cite{BMW:2014pzb,Chang:2018uxx,Berkowitz:2018gqe,Hall:2025ytt}.  However, predicting NN interactions observed in nature remains an outstanding theoretical challenge.

This is due in part to the fact that the full pipeline of calculations required to compute the NN (and NNN) interactions from QCD have not been fully ported to utilize GPUs.
As we describe in the next section, we have implemented well known algorithms, that are necessary for these calculations, to make efficient use of GPUs, achieving a {\it game changing} innovation for this field and a speed up of \su{} times over previous state of the art.
We have applied this new implementation to the calculation of the NN interaction with quark mass parameters ($m_q$) that are equal to those of nature as well as about twice as heavy, and the first time a calculation with such a small quark mass has been performed.  This will allow us to begin to address questions directly from QCD, including: {\it How sensitive is the composition of the universe to small changes in the fundamental parameters of QCD?}
{\it What are the strengths of the nuclear reactions that power the stars and make life possible?}

\section{\label{sec:curent_sa} Current state-of-the-art}

%
%

In recent years, LQCD computations have finally been able to reliably study nucleon-nucleon systems, as well as evaluate the masses and decay widths of unstable hadron resonances, such as the $\rho$ and $\Lambda(1405)$ resonances.  This remarkable achievement requires first computing the finite-volume energies of the multi-hadron states, into which a resonance decays, using Markov-chain Monte Carlo path integration.
{Next, the functional dependence of the scattering amplitudes on the energy of the system is constrained using the well-known finite-volume formalism~\cite{Luscher:1986pf,Luscher:1990ux}. In particular, the scattering amplitude is parametrized by a few parameters and their values are determined by standard statistical inference tools. The properties of unstable hadrons follow from the scattering amplitudes.}

A key step in such computations is the evaluation of the energies of the multi-hadron states. These energies are extracted from Monte Carlo estimates of temporal correlations involving quantum field operators that create the states of interest.  To evaluate such correlators, quark propagators from a variety of source sites on the lattice must be knitted together.  The quark propagators themselves are the inverses of an exceedingly large matrix, 
but fortunately, only the matrix products of the inverse and the various source sites is needed.  
These correlation functions are described by a sum of exponentials,
\begin{equation}
C(t) = \sum_{n=0}^N |z_n|^2 e^{-E_n t}\, ,
\label{eq:correlatortimedependence}
\end{equation}
where the time-independent ``overlap'' factors, $z_n$ measure the coupling between the vacuum and a given state $n$ with the creation and annihilation operators used.
The time-dependence of the correlation functions is captured with the decaying exponentials that depend upon the energy of the states created.  For a given correlation function, we are typically only interested in the lowest energy state, $E_0$, and there is usually a gap to the excited state energies, $E_n-E_0 > 0$, and so for sufficiently large times $t$, we can extract the spectrum of interest.

For correlators involving stable hadrons, translational invariance can be used to limit the number of source sites to a small few.  For multi-hadron operators, such as two-nucleon ones, all spatial sites on a source time slice must be used, significantly increasing the numerical cost, and therefore, reliable estimates of such correlations involving multi-hadron operators were not feasible to attain until recently.
LQCD calculations of NN are a notoriously difficult exascale grand challenge problem~\cite{np_exa} for several additional reasons.  The two most challenging aspects of the problem, beyond the cost mentioned above, are:
i) The physics of interest residing in the interaction energy between the nucleons, which as discussed above, is of order 0.1\% of the total energy of the system, and must be determined precisely and accurately through calculations of the total energy.
ii) At the same time, LQCD calculations suffer from an exponentially bad signal to noise (S/N) problem~\cite{Lepage:1989hd,Parisi:1983ae} where a system of $A$ nucleons has a S/N that degrades in time, $t$, as 
\begin{equation}
\frac{\textrm{Signal}}{\textrm{Noise}}
	\sim \sqrt{N_{sample}}\ e^{-A(m_N - \frac{3}{2}m_\pi)t}\, .
\end{equation}
with $m_N$ the nucleon mass and  $m_\pi$ the pion mass (the pion, $\pi$,  is the lightest particle in QCD). In nature $m_N \approx 938 $ MeV and 
$m_\pi \approx 140$ MeV.
These masses emerge from the input quark masses and QCD dynamics and so in numerical LQCD computations, we can explore many different values in our quest to understand how the NN interaction changes and becomes fine tuned.
For all interesting values, this S/N problem means that it is not possible to compute the correlation function for long enough time $t$ to extract the lowest state in Equation~\ref{eq:correlatortimedependence}, even with exascale computing.  Rather, an improved method is required.

A novel quark-field smearing scheme, known as Laplacian Heaviside (LapH)~\cite{Peardon:2009gh}, has now made such reliable estimates feasible. LapH quark smearing projects the quark propagators into a smaller subspace spanned by various eigenvectors of the gauge-covariant Laplacian, allowing the use of all spatial sites on a source time slice in a feasible manner.
The utilization of the LapH method ameliorates the S/N issue by enabling a determination of the spectrum early in time (thus small $t$ above), before the S/N problem becomes too severe. 
However, the method is sufficiently expensive for NN calculations that it was only a few years ago that it was first applied to NN systems, at a pion mass that is much heavier than nature~\cite{Horz:2020zvv}, and on CPU architectures using a stochastic variant of the LapH method~\cite{Morningstar:2011ka}.
In order to carry out such LQCD calculations of NN systems, at pions masses that are close to the physical pion mass, it is essential to port the entire workflow of the calculations to utilize GPUs.

\smallskip

\section{\label{sec:innovations} Innovations}

\subsection{LapH}

Getting such calculations to run efficiently on a variety of CPU-GPU
architectures is a difficult task.  Much effort has gone into developing
a library known as \quda{}~\cite{quda:software} to facilitate such calculations.  Before this work,
\quda{} was mainly developed for carrying out the matrix inversions needed
for quark propagation.  This work is focused on developing efficient
GPU code in \quda{} for evaluating the eigenvectors of the three-dimensional
gauge-covariant Laplacian needed for LapH quark smearing, as well as
on implementing driver software with an efficient environment, known as \qudalaph, to carry out the sequence of tasks needed for our physics
program.

\subsection{\Qudalaph}
\label{sec:qudalaph}
We have developed a software suite known as \qudalaph{}~\cite{quda_laph} which drives the user requested tasks using \quda{} as a library.  \Qudalaph{} is mainly CPU code written in C++ with MPI and OpenMP threads which handles reading the input XML that specifies the tasks to perform, reading data from file, writing data to file, and calling routines in \quda{} to carry out the tasks.  Persistent
data needed for more than one task are handled with a singleton class.  For the most part, \quda{} handles copying and/or moving data from the host to the device and back.  \Qudalaph{} is coded very close to the low-level optimized kernels of \quda, avoiding too many conversion routines and unnecessary data copying and moving.  An additional Dirac spin basis convenient for our calculations was also introduced.

\subsection{\quda}
\label{sec:quda}
\quda{} is an open-source library to provide GPU acceleration for LQCD computations~\cite{quda:software,Clark:2009wm,Babich:2011np}.  With respect to this work, the relevant features of \quda{} include
\begin{itemize}
\item Aggressive use of {\bf mixed-precision} to accelerate the bandwidth-bound sparse computations.
\item Initially a CUDA-only library, it is now {\bf portable and runs on CUDA, HIP and SYCL} platforms through an extensive parallelism abstraction and separation of data order from computation through the use of opaque accessors.
\item The workhorse of any LQCD computation is the Dirac matrix iterative linear solver, and \quda{} features a {\bf class-leading implementation of the adaptive geometric multi-grid solver}~\cite{Clark:2016rdz}.
\item Portability across platforms, and indeed across generations of GPU architecture from the same vendors is achieved through the use of {\bf aggressive autotuning} of GPU parameters (threads per block, grid size, shared memory size) as well as communication policies (DMA bulk copies versus fine-grained remote write).
\end{itemize}

While \quda{} could offload the computationally demanding Dirac matrix linear solves to GPUs, no GPU-based implementation was previously available to evaluate the LapH eigenvectors nor to carry out projections of the quark propagators onto these eigenvectors.  This was a very serious
bottleneck for LQCD studies involving LapH quark-field smearing.
The LapH eigenvectors are needed as sources for the quark propagator
computations, for projecting the resulting propagators onto the
LapH eigenvectors, and for forming so-called quark-antiquark doublets and
three-quark triplets to facilitate the Wick contractions involving both
meson and baryon sources and sinks in temporal correlators. With LapH
smeared quark fields, computing the LapH eigenvectors or reading them from
file must be done at the start of every run.  Previous CPU-based code to
compute the LapH eigenvectors was uncomfortably slow, requiring a comparable 
time as for a large number of linear solves.  This necessitated that the eigenvectors would be computed offline and saved to disk, requiring hundreds to thousands of files of several TB each.  This process is both cumbersome as well as inefficient: while storing and reusing eigenvectors was faster than recomputing inline, the overhead of loading the eigenvectors still consumed a significant amount of the runtime.  A new solution was required to avoid the file i/o overhead, and make for a more convenient workflow.

\subsubsection{Scalable Lanczos Eigensolver \label{sec:lanczos}}

To remove the bottleneck of offline eigenvector generation and storage, we have implemented a highly-scalable batched Lanczos eigensolver in \quda.  No output to disk is required anymore, and typically in
a few minutes, the eigenvectors are computed and available, taking at least a factor of two less time than reading them from disk.  The GPU-based
code to carry out the projections onto the LapH eigenvectors is also now
an order of magnitude faster than before.

Given that we are concerned with the 3-d (spatial) Laplace operator, the temporal extent of the lattice dimension is an independent batching dimension upon which no communication occurs.  This provides an avenue to reduce communication, both neighborhood halo communication and the domain size of global reductions.
\begin{itemize}
    \item For halo communication, we optimize our 4-d domain decomposition to prefer splitting spatial dimensions (X/Y/Z) locally within a node.  This ensures that the eigensolver gets the most benefit from fast intra-node communication over NVLink / InfiniFabric.
    \item We maximize the scalability of global reductions required in the Lanczos solver through the use of MPI communicators: for each 3-d process sub-domain we introduce a communicator and run the batched eigensolver restricted to that communicator.  For example, given a global extent of the lattice of \(L_x \times L_y \times L_z \times L_t\), and a process topology given by \(N_x \times N_y \times N_z \times N_t\), we will have \(N_t\) sub-communicators, each with a process decomposition \(N_x \times N_y \times N_z\), each running a batched Lanczos eigensolver with local batch size \(L_t / N_t\).
\end{itemize}

\subsubsection{Eigenvector Projection \label{sec:projection}}

An important stage in the pipeline is the projection of the propagators onto the LapH eigenvectors; this having previously being done on CPUs, however, this is time consuming given the number of inner products required over the set of basis vectors.  While computationally straightforward, a simple port to GPUs is not possible given the limited GPU memory, coupled with the projection space of the number of eigenvectors multiplied by the number of solution vectors.  As a solution we employed a tiled approach to the problem: we perform a two-dimensional tiling over the eigenspace and the list of solution vectors, and work on a tile of the problem, computing the set of inner products corresponding to that tile, before moving on to the next tile.
This allows for the amortization of the CPU to GPU transfers.

\subsubsection{Batched Linear Solves \label{sec:batched}}

New to this work is the inclusion of the recently developed batched linear solver: batching linear solves recasts a matrix-vector problem into a matrix-matrix problem. This transformation is beneficial for three reasons
\begin{enumerate}
    \item Batching increases the temporal locality of the problem, since the stencil coefficient in the linear system are reused across the batch.  This reduces the bandwidth dependence, increasing the utilization.
    \item Batching trivially increases the parallelism of the problem, allowing for better saturation of GPU resources.  This is especially important in the context of multi-grid, where the coarse grids have limited parallelism.
    \item The improved utilization, coupled with decreased memory traffic increases the energy efficiency of the computation.  This is of increasing importance as we approach the end of a transistor energy efficiency scaling with the slow down of Moore's Law.  
\end{enumerate}

Through the use of batching, we found up to a 3x fold improvement in solver throughput on the Alps, Jupiter, and Perlmutter clusters. However, due to limited access, our results on El Capitan and Frontier do not include the use of batching yet. We are currently implementing this and it will be available shortly. We shall update our results to include the benefit of batching on all machines and report in a future update if it is offered to us.

\section{\label{sec:perf_meas_method} Performance Measurement Method}

In the last few years LQCD has made great advances in our
understanding on how to compute very complicated and computationally
challenging quantities, such as nucleon-nucleon interactions, that are central to our understanding of nature. Our group has been instrumental in entering this new era with advanced physics methods, algorithms and supercomputing codes as described in the previous sections. 
These advances are a perfect match to the new exascale-class supercomputer architectures and resources.

In the past it was not possible to fit the full lattice in a small
number of nodes. Instead, it had to be spread out on a significant
number of nodes exposing the problem to the slow network
communications compared to node computations. This lead to slower
strong scaling that restricted the computational capability. The new
exascale class supercomputers have extremely powerful nodes with large
local memory but relatively slow networks. This has resulted to even smaller
communication to computation speed ratios signaling the need for a new
paradigm of computing our problem. We use the powerful nodes and large
local memory to fit a 
basic unit of computation
on a small number of nodes (8 nodes for El Capitan) while we use the other parameters of our problem 
to scale our computations to a large number of nodes.

A full LQCD calculation requires several large scale Monte Carlo integrals to be carried out.  First, because the dynamical parameters, such as the mass of the nucleon and the two-nucleon interaction energy, are dynamically generated from QCD through non-perturbative interactions, we do not know {\it a priori} what input values of the quark masses to use.  Therefore, several calculations at different values must be used in order to extrapolate, or ideally interpolate, to the values that reproduce the observed nucleon masses of nature.  Second, we must carry out the computation at three or more values of the lattice spacing such that we can systematically remove discretization errors that otherwise distort the results from those in nature.  As with the quark mass parameters, because the theory is non-perturbative, the scale of the lattice spacing is not {\it a priori} known and it must be inferred by computing some reference mass scale.  Third, an extrapolation to the infinite volume must also be performed, which requires performing the calculation at different values of the volume.

Each choice of quark mass, lattice spacing and volume is referred to as an {\it ensemble}.  For each ensemble, in order to precisely estimate the integral with Monte Carlo methods requires us to perform the calculation on roughly 1000 independent snapshots of the QCD vacuum, referred to as {\it configurations} which are generated in a previous step of LQCD calculations to those needed to compute the two-nucleon interactions.  The calculation of the two-nucleon correlation functions on each configuration is an independent computation that must be performed.  Further, with the LapH method, the calculation on each time-slice of each configuration is independent.  For the ensembles used in this work, the dimensionless temporal extend is given by $N_t=192$, and therefore, for a single ensemble of 1000 configurations, a full calculation requires 192,000 independent MPI tasks to be performed.  In this sense, the new exascale computers with very powerful nodes relative to the interconnect, is a perfect match for our science problem.

For example, to fill El Capitan, which has $\sim11,000$ nodes, we can compute 7 configurations of all 192 times slices as each sub-task requires 8 nodes, which requires 10,752 nodes.  The only common resource between these tasks is the
file and queuing systems that exhibit minimal to no overlap. In this
new era this is the best way to take full advantage of the new
supercomputers for our problem.

For this reason, the natural metric to measure our software is the time to solution for the entire calculation of an ensemble.  In some ways, this is similar to strong scaling.
In traditional HPC strong scaling measurements, the problem size is fixed while the number of nodes used is increased.
Using this as an analogy, we can consider a complete calculation on one ensemble as a fixed problem, and measure how efficiently this problem can be computed as the number of nodes is increased.  Beyond the minimal number of nodes required for an independent sub-task this computation does not require further network communications aside from the shared file-system and queuing system.  While this is not a strong-scaling test in the traditional sense, it is the
most meaningful metric for our science problem as it measures the real time required to complete a calculation as we scale up in the number of nodes.
We define this as ``resource scaling''.

For an explicit measurement with our problem, we take a fixed global problem that has a dimensionless volume of $V=64^3\times192$ where the last dimension is the number of time slices.  As mentioned above, we need to perform a calculation on each of the 192 time slices.  Further, we want to carry out the calculation on 1000 configurations.  We define a single, independent MPI job on 8-nodes as a sub-task, while a job submitted to the queueing system as a job.
Each individual sub-task requires 8 nodes, and so a 16 node job will simultaneously run 2 sub-tasks for example.
We measure the wall clock time to carry out such calculations as we increase the number of 8-node sub-tasks performed in a single job submitted to the queueing system.  Flux~\cite{AHN2020202} was used to manage the sub-tasks in the subsequently larger jobs submitted.  The wall clock times measured include both I/O as well as the startup time the queueing system takes to start all sub-tasks.
These timings therefore represent the full resources needed to run a given sub-set of the full calculation.

In addition, we measure traditional strong scaling of wall-clock time from 6 to 128 El Capitan nodes including I/O. The performance gain significantly saturates after 8 nodes, and therefore we choose 8 nodes to be our ``unit'' size.

For weak scaling we measured performance as the wall-clock time it
took for our application to run including I/O on El Capitan and Alps. 
The wall clock time was measured by a timer embedded in our code.

The kernel of our problem is a mixed-precision multi-grid 
solver of a sparse matrix as described above. We measure its performance with an embedded timer and give the results in the same plots as for the full application.

We define the previous state-of-the-art timings as the ones we measured during our previous physics calculation of a very similar problem with the same sub-task problem size. The physics results were published in~\cite{Bulava:2022vpq} and were computed on the CPU-only Frontera supercomputer. The LapH method described above 
previously %
was 
only %
possible to implement on CPUs but not on CPU-GPU platforms. Our new work, as described above, has enabled the implementation of LapH on CPU-GPU architectures. We compare the previous state-of-the-art with the timings we obtain on several of the new exascale-class supercomputers; Alps \cite{ALPS}, El Capitan \cite{ElCapitan}, Frontier \cite{Frontier}, Jupiter \cite{Jupiter}, and Perlmutter \cite{Perlmutter}.

\section{\label{sec:perf_results} Performance Results}

We performed scaling studies during a 6 hour dedicated access to the full LLNL El Capitan supercomputer as well as a dedicated run on a large fraction of the Alps supercomputer (2304 of the 4-socket Grace-Hopper nodes). At the moment of this writing El Capitan is the fastest supercomputer in the world with peak speed of 2.79 exaFLOPS and more than 11,000 nodes. At the time of the study a few nodes were not available but we used a maximum number of 10,752 nodes. 
It is remarkable that scaling runs of a few hours produced a good fraction of the data needed for our nucleon-nucleon scattering project. This is a testament of the performance of our full methods, physics, algorithms, and implementation, as described in this paper. After performing our scaling runs, El Capitan switched to LLNL classified access. We now have access to the sister unclassified system Tuolumne. 

\begin{figure}[t]
\includegraphics[width=1.0\linewidth]{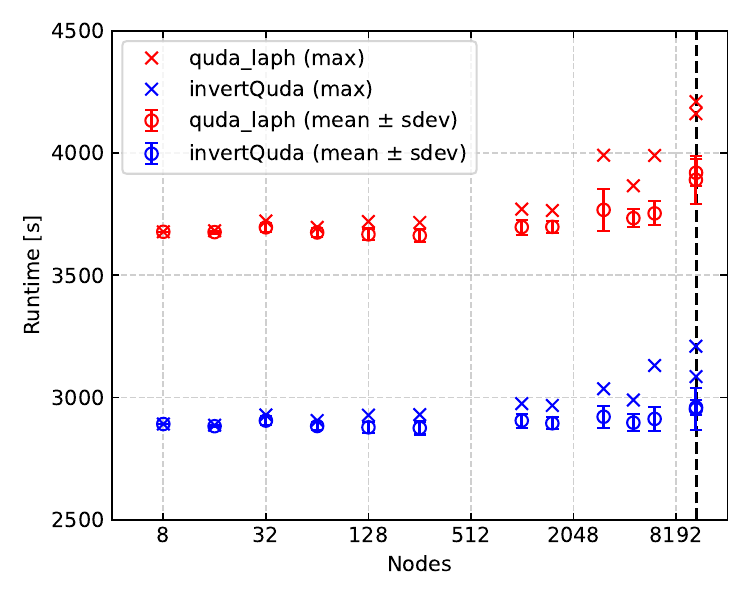}
\caption{We present the wall-clock time of full production calculations in a weak scaling study, including I/O, on El Capitan. The black dashed vertical line indicates 10,752 nodes.}
\label{fig:weak_scaling}
\end{figure}

In Figure \ref{fig:weak_scaling} we present our weak scaling results on El Capitan.
The wall-clock runtime, excluding job scheduler overhead, is measured for each sub-job, and the maximum (cross), mean, and standard deviation (circle with errorbar) are plotted. \Qudalaph{} (in red) is the software described in Sec.~\ref{sec:qudalaph} which handles all the calculations and file I/O. ``invertQuda'' (in blue) refers to the dominant kernel of \qudalaph{} (the linear solver). 
The black dashed vertical line indicates 10,752 nodes, representing the total number of El Capitan nodes available at the time. Near-perfect linear scaling can be observed.

\begin{figure}[t]
\includegraphics[width=1.0\linewidth]{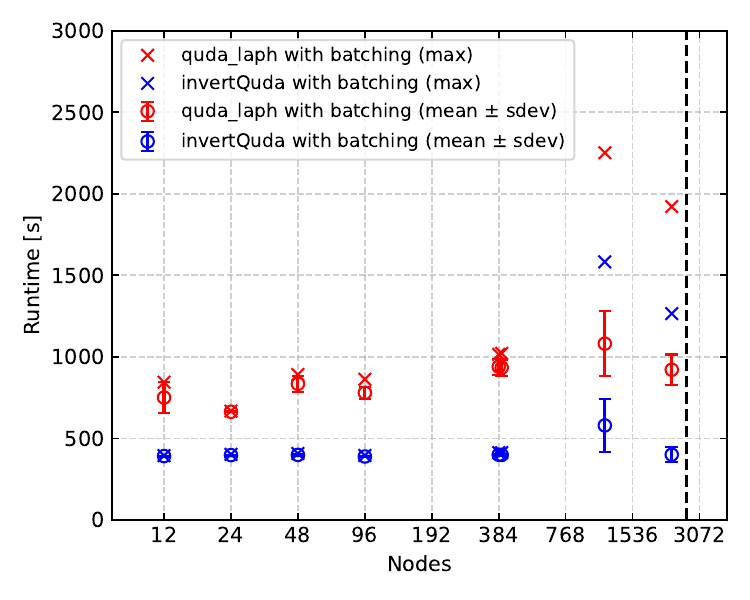}
\caption{We present the wall-clock time of full production calculations in a weak scaling study, including I/O, on Alps utilizing the batched linear solver optimizations }
\label{fig:weak_scaling_alps}
\end{figure}

In Figure \ref{fig:weak_scaling_alps} we present our weak scaling results from the Alps supercomputer. These results use our latest code improvement, ``quda\_laph with batching", described in section \ref{sec:batched}. As of this writing, we have run our new code to measure weak scaling only on Alps. We are working to do the same for El Capitan, Frontier, Jupiter, and Perlmutter supercomputers. Again, near-perfect weak scaling is observed.

\begin{figure}[t]
 \includegraphics[width=1.0\linewidth]{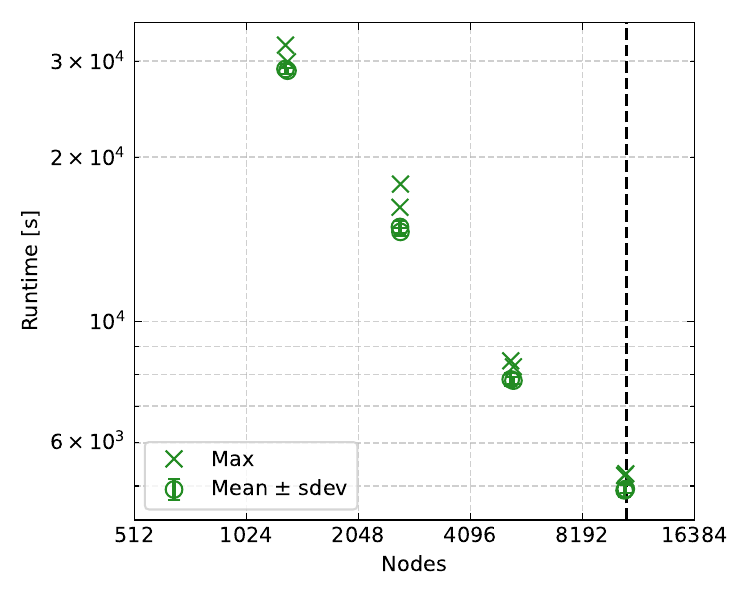}  
\caption{``Resource scaling'' of the global problem on El Capitan.  This includes the full resource cost, such as I/O and job scheduler overhead. The black dashed vertical line indicates 10,752 nodes.}
\label{fig:job_scaling}
\end{figure}

We present our results for ``resource scaling'' in Figure~\ref{fig:job_scaling}, where we plot the total wall-clock time on El Capitan for our complete production run for various numbers of nodes. Scaling was performed using Flux \cite{AHN2020202}, a resource management and job scheduling framework. Note that we measure the full batch job timing, i.e. including I/O and job scheduler overhead of the batch job that launches all sub-jobs. 
Furthermore, we indicate the maximum (cross), mean and standard deviation (circle with error bar) of the measured times. 
The black dashed vertical line indicates 10,752 nodes, again representing the total number of El Capitan nodes available at the time. Nearly perfect linear scaling can be seen: the total runtime reduces proportionally with increasing number of nodes while our global problem is held fixed.
This is a non-trivial measure as batches of up to 192 sub-tasks in each job were reading the same multi-gigabyte input file required at the beginning of each calculation.  Further, it demonstrates the ability of Flux to simultaneously launch all sub-tasks without significant latency.

\begin{figure}
  \includegraphics[width=1.0\linewidth]{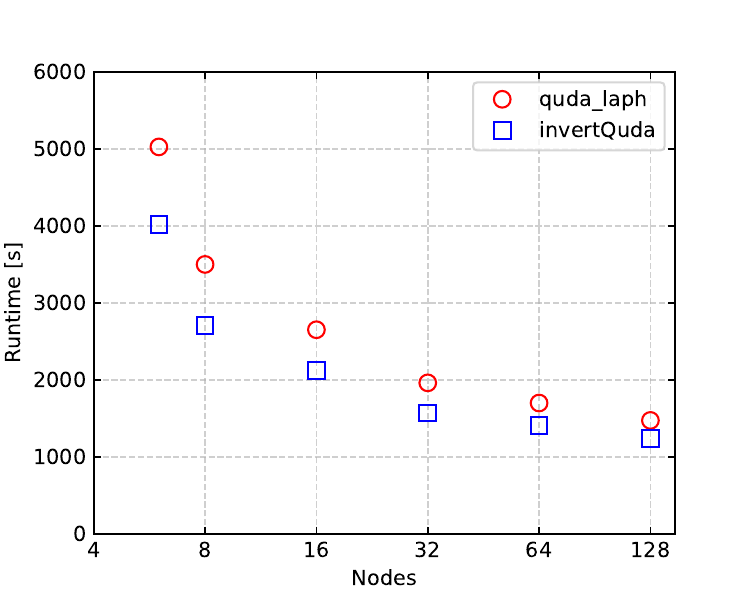}
\caption{Strong scaling analysis (including I/O), indicating that our choice of a sub-job size of 8 nodes is optimal.}
\label{fig:strong_scaling}
\end{figure}

Subsequently, in Figure \ref{fig:strong_scaling} we present our results of 
a strong scaling analysis on El Capitan using one of the sub-tasks (fixed problem size). These timings include I/O, but exclude job scheduler overhead, and were obtained on El Capitan with up to 128 nodes. The total number of MPI processes (total number of GPUs) for each job is $4\times $nodes. 
Using more nodes would not have been meaningful for two reasons. First, because at 128 nodes we clearly had left the region of linear scaling. Second, because we were searching for the optimum from where on to start using resource scaling, aiming for a runtime of one hour for our global problem. From these results, it is obvious that an optimal choice is to use 8 nodes, as we have stated above. 

\begin{figure}[t]
  \includegraphics[width=1.0\linewidth]{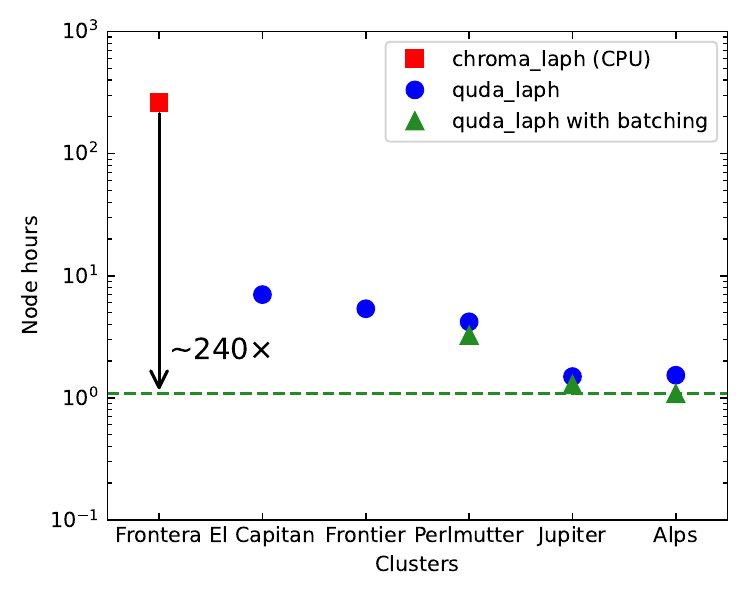}
\caption{Comparison of the previous state-of-the-art (red) with some of today's exascale-class supercomputers (blue/green). For our code we measure up to a factor of \su{} speedup. Note that
the node-hour measurements are not suitable as a direct comparative benchmark between the new supercomputers due to variations in the job parameters from one machine to another, such as the number of nodes and/or ranks used.
}
\label{fig:comparison}
\end{figure}

A comparison of the previous state of the art (red) \cite{Bulava:2022vpq} with some of today's exascale-class supercomputers (blue/green) is shown in Figure \ref{fig:comparison}. 
This comparison uses the same fixed sub-task size described above: a calculation that requires the full dimensionless volume of $V=64^3\times192$ for the quark-propagators, but the creation of a ``source'' for the propagators in the LapH sub-space on one out of the 192 time slices.
We measure
the wall clock time and multiply with the number of nodes. Here, ``chroma\_laph (CPU)" refers to the the previous state-of-the-art on the CPU only Frontera architecture. Similarly, ``quda\_laph" refers to the GPU accelerated method described in Sec.~\ref{sec:qudalaph}. Finally, ``quda\_laph with batching" refers to the our latest code improvement as described in section \ref{sec:batched}. As of this writing, timing for it is included for the NVIDIA-powered Alps, Jupiter, and Perlmutter supercomputers and will be implemented shortly in the AMD-powered El Capitan and Frontier supercomputers. For our code we measure up to a factor of \su{} speedup. 
Note that the node-hour
measurements are not suitable as a direct comparative benchmark between
the new supercomputers due to variations in the job parameters from one
machine to another, such as the number of nodes and/or ranks used.

In addition to the scaling studies reported in detail, which were performed on the ensemble with a quark mass about twice as heavy as nature, we also ran computations on an ensemble with the physical value of the quark mass.  This ensemble has a dimensionless volume of $96^3\times192$ as compared to the $64^3\times192$ used in our studies, making some aspects of the computation significantly more expensive as well as suffering from a more severe signal-to-noise problem, thus requiring exponentially more statistics for precise results.  The speed up on this ensemble is even larger than for the one presented.  In order to maximize the science output of our scaling runs, we focused on the less expensive ensemble for which we anticipate obtaining enough results for a publication.

\section{\label{sec:impl_results} Implications}

Our ability to understand the emergence of nuclear physics directly from QCD and the Standard Model remains a notoriously difficult exascale challenge.
Making such a quantitative connection has important implications both for our understanding of basic nuclear physics processes with controlled uncertainty, such as understanding the fine-tunings present in nature, as well as 
being required for precision tests of the Standard Model and the search for new physics.  In order to achieve these goals, it is essential to enable the full lattice QCD workflow for computing nuclear physics processes, such as two-nucleon scattering, to make efficient use of the heterogeneous CPU-GPU architectures that are prevalent in the exascale era.

The work presented here represents the first highly optimized implementation LapH method with GPUs for computing quark propagators needed for two-nucleon calculations, which is implemented in the open-source \quda{} library.
This work also represents the first demonstration of the batched linear solver recently implemented in \quda{} in a publication.
With these advances, we demonstrate for the first time, the ability to efficiently utilize CPU-GPU architectures to carry out two-nucleon calculations at pion masses very near the physical value, obtaining an \su{} speed up over previous state-of-the-art CPU implementations.

We compare our current results to previous CPU implementations because, prior to the GPU implementation of the \textit{scalable Lanczos Eigensolver} and \textit{Eigenvector Projection}, it was not practically feasible to run the computations, despite having highly optimized linear solvers for the Dirac matrix.  Either the computations would require very significant I/O burden in rate and total size, or, the eigenvectors would be computed on the CPU components of heterogeneous nodes then passed to the GPUs for the linear Dirac matrix solves.  Both approaches were sufficiently prohibitive as not to be utilizes.  The GPU implementation of the eigenvector solver and projector, Secs.~\ref{sec:lanczos} and \ref{sec:projection}, is a game changer for these types of computations, allowing us to take full advantage of the efficient GPU implementation of Dirac matrix linear solves implemented in \quda, including the newly described in Sec.~\ref{sec:batched}.

This represents a significant milestone on applications of lattice QCD to nuclear physics.
We are now able to perform a computation that previously required one year in a few days.
This disruptive improvement enables a new era of scientific research in the quest to understand the exciting mysteries of nature at the most fundamental level of nuclear and sub-nuclear (QCD) physics and will contribute to our understanding of the processes that construct our cosmos and make life possible.

As the new generation of supercomputers has powerful nodes with large memory but relatively slow inter-node networks, the best way to map our problem onto the machine is different than previous generations, as described in this work. Perhaps this natural change of paradigm may result into a continuing line of supercomputing architectures in the future where the node (or small cluster of nodes) has very large computational speed and memory size, while the network is locally extremely fast and can keep up with computation while it is slow for distant nodes.

Our physics methods, algorithms, and code implementations are publicly available as cited in this work~\cite{quda:software,quda_laph} and can be used by a wide range of researchers in this field or inspire advances in other fields that share similar mathematical problems. We hope that our work will also contribute to expanding the user base of the new extraordinarily powerful machines.

\section*{Acknowledgement}
This work was performed under the auspices of the U.S. Department of Energy (DOE)
by Lawrence Livermore National Laboratory under Contract DE-AC52-07NA27344
(ASM, SP, PMV). ASM, SP, and PMV acknowledge the support from the NNSA ASC COSMON project and would like to thank the LLNL ASC Program Office and
Scott Futral of LLNL for their support and early access to LLNL’s exascale systems, Tuolumne and El Capitan. CM acknowledges support from the NSF under award PHY-2209167 and OAC-2311430.
AWL acknowledges support from the DOE Office of Science, Office of Nuclear Physics under contract number DE-AC02-05CH11231 and from the NSF under award number OAC-2311431.
SK is partially supported by the DFG through project no. 460248186 (PUNCH4NFDI), project no. 511713970 (NuMeriQS), by the MKW NRW under funding code NW21-024-A, and by the Helmholtz Association through the AIDAS laboratory.
The work of FRL was supported in part by the Platform for Advanced Scientific Computing (PASC) project ``ALPENGLUE''. Part of the computations used a grant from the Swiss National Supercomputing Centre (CSCS) under project ID lp53 
and g193 %
on Alps.
This work is supported in part by the Neutrino Theory Network Program Grant
DE-AC02-07CHI11359, and U.S. Department of Energy Award DE-SC0020250 (ASM). This research used resources of the Oak Ridge Leadership Computing Facility at the Oak Ridge National Laboratory, which is supported by the Office of Science of the U.S. Department of Energy under Contract No. DE-AC05-00OR22725 (HMC).

\bibliographystyle{IEEEtran}
\bibliography{GBbib}

\end{document}